\documentclass[a4paper]{jpconf}
\usepackage{graphicx}
\begin{document}

\title{Clustering effects induced by light nuclei}

\author{C Beck}

\address{D\'epartement de Recherches Subatomiques, Institut Pluridisciplinaire 
Hubert Curien, IN$_{2}$P$_{3}$-CNRS and Universit\'e de Strasbourg - 23, rue 
du Loess BP 28, F-67037 Strasbourg Cedex 2, France}

\ead{christian.beck@iphc.cnrs.fr}

\begin{abstract}
Since the pioneering discovery of $^{12}$C+$^{12}$C 
molecular resonances half a century ago, a great deal of research work has been undertaken in 
the $\alpha$-clustering study. Our knowledge on  
the physics of nuclear molecules has increased considerably and nuclear 
clustering remained one of the most fruitful domains of nuclear physics, 
facing some of the greatest challenges and opportunities in the years ahead. 
In this work, the occurrence of ``exotic'' shapes in light $N$=$Z$ $\alpha$-like 
nuclei is investigated. Various approaches of the superdeformed and hyperdeformed 
bands associated with quasimolecular resonant structures are presented. 
Clustering aspects are also briefly discussed for light nuclei with
neutron excess trough very recent results on neutron-rich Oxygen isotopes.
\end{abstract}

\section{Introduction}

The search for resonant structures in the excitation functions for various 
combinations of light $\alpha$-cluster ($N$=$Z$) nuclei in the energy regime 
from the Coulomb barrier up to regions with excitation energies of $E_{x}$=20$-$50~MeV 
remains a subject of contemporary debate~\cite{Greiner95,Beck94}. These 
resonances have been interpreted in terms of nuclear molecules~\cite{Greiner95}. 
The question of how quasimolecular resonances may reflect continuous transitions
from scattering states in the ion-ion potential to true cluster states in the 
compound systems is still unresolved \cite{Greiner95,Beck94}. In many cases, these 
resonant structures have been associated with strongly-deformed shapes and with 
alpha-clustering phenomena \cite{Freer07,Horiuchi10}, predicted from the 
Nilsson-Strutinsky approach, the cranked $\alpha$-cluster model~\cite{Freer07}, or 
other mean-field calculations~\cite{Horiuchi10,Gupta10}. In light $\alpha$-like 
nuclei clustering is observed as a general phenomenon at high excitation energy 
close to the $\alpha$-decay thresholds \cite{Freer07,Oertzen06}. This exotic 
behavior has been perfectly illustrated by the famous "Ikeda"-diagram for $N$=$Z$ 
nuclei in 1968 \cite{Ikeda}, which has been recently extended by von Oertzen 
\cite{Oertzen01} for neutron-rich nuclei, as shown in the left panel of Fig.1.
Clustering is a general phenomenon \cite{Milin} not only observed in light
neutron-rich nuclei \cite{Kanada10}, but also in halo nuclei such as $^{11}$Li 
\cite{Ikeda10} or $^{14}$Be, for instance \cite{Nakamura11}. The problem of 
cluster formation has also been treated extensively for very heavy systems by R.G. Gupta \cite{Gupta10} 
and by Zagrebaev and W. Greiner \cite{Zagrebaev10} where giant 
molecules and the collinear ternary fission may exist \cite{Kamanin}.  

\newpage

\begin{figure}
\includegraphics[scale= 0.72]{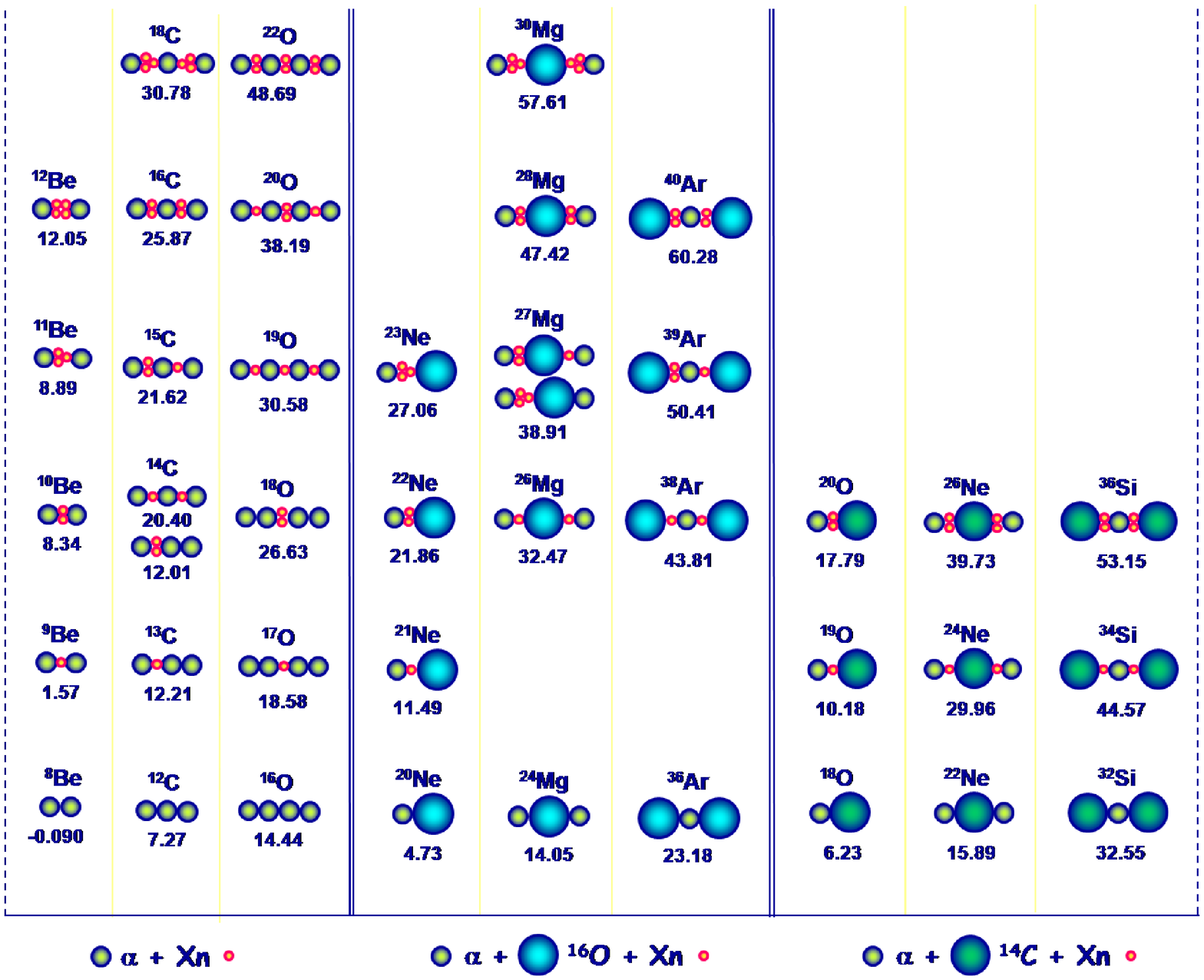}
\caption{\label{label}Schematic illustration of the structures of molecular
shape isomers in light neutron-rich isotopes of nuclei consisting
of $\alpha$-particles, $^{16}$O- and $^{14}$C-clusters plus some
covalently bound neutrons (Xn means X neutrons) \cite{Milin}. The so called "Extended 
Ikeda-Diagram" \cite{Oertzen01} with $\alpha$-particles (left panel) and 
$^{16}$O-cores (middle panel) can be generalized to $^{14}$C-cluster cores 
(right panel). The lowest line of each configuration corresponds to parts
of the original Ikeda diagram \cite{Ikeda}. However, because of its deformation,
the $^{12}$C nucleus is not included, as it was earlier \cite{Ikeda}.
Threshold energies (in MeV) are given for the relevant 
decompositions.}
\label{fig:1}
\end{figure}

\section{Alpha clustering, deformations and alpha condensates}
\label{sec:1}

The real link between superdeformation (SD), nuclear molecules and alpha 
clustering \cite{Horiuchi10,Beck04a,Beck04b,Cseh09} is of 
particular interest, since nuclear shapes with major-to-minor axis ratios of 
2:1 have the typical ellipsoidal elongation  for light nuclei (with quadrupole deformation 
parameter $\beta_2$ $\approx$ 0.6). Furthermore, the structure 
of possible octupole-unstable 3:1 nuclear shapes (hyperdeformation (HD) with $\beta_2$ $\approx$ 
1.0) has also been 
discussed for actinide nuclei \cite{Cseh09} in terms of clustering phenomena. Typical examples for 
possible relationship between quasimolecular bands and extremely deformed (SD/HD) 
shapes have been widely discussed in the literature for $A = 20-60$ $\alpha$-conjugate 
$N$=$Z$ nuclei, such as $^{28}$Si 
\cite{Taniguchi09,Ichikawa11,Jenkins12,Jenkins13,Darai12}, $^{32}$S 
\cite{Horiuchi10,Kimura04,Lonnroth10,Chandana10,Ichikawa11}, 
$^{36}$Ar \cite{Cseh09,Beck08a,Svensson00,Sciani09,Beck09}, $^{40}$Ca 
\cite{Ideguchi01,Rousseau02,Taniguchi07,Norrby10}, $^{44}$Ti 
\cite{Horiuchi10,Leary00,Fukada09}, $^{48}$Cr \cite{Salsac08} and $^{56}$Ni 
\cite{Nouicer99,Rudolph99,Beck01,Bhattacharya02}.

\newpage

In fact, highly deformed shapes and SD rotational bands have been 
discovered in several light $\alpha$-conjugate ($N$=$Z$) nuclei, such as $^{36}$Ar
and $^{40}$Ca by using $\gamma$-ray spectroscopy techniques 
\cite{Svensson00,Ideguchi01}. In particular, the extremely deformed rotational
bands in $^{36}$Ar \cite{Svensson00} (shown as crosses in Fig.~2) might be 
comparable in shape to the quasimolecular bands observed in both $^{12}$C+$^{24}$Mg 
\cite{Sciani09,Cindro79,Mermaz84,Pocanic85} (shown as open triangles in Fig.~2)
and $^{16}$O+$^{20}$Ne \cite{Shimizu82,Gai84} (shown as full
rectangles) reactions. Ternary clusterizations are also predicted 
theoretically, but were not found experimentally in $^{36}$Ar so far 
\cite{Beck09}. On the other hand, ternary fission of $^{56}$Ni -- related to its hyperdeformed 
shapes -- was identified from out-of-plane angular correlations 
measured in the $^{32}$S+$^{24}$Mg reaction with the Binary Reaction 
Spectrometer (BRS) at the {\sc Vivitron} Tandem facility of the IPHC, Strasbourg 
\cite{Oertzen08}. This finding \cite{Oertzen08} is not limited to light 
$N$=$Z$ compound nuclei, true ternary fission \cite{Zagrebaev10,Kamanin,Pyatkov10}
can also occur for very heavy \cite{Kamanin,Pyatkov10} and superheavy 
\cite{Zagrebaev10b} nuclei.

\begin{figure}[h]
\begin{minipage}{18pc}
\includegraphics[width=18pc]{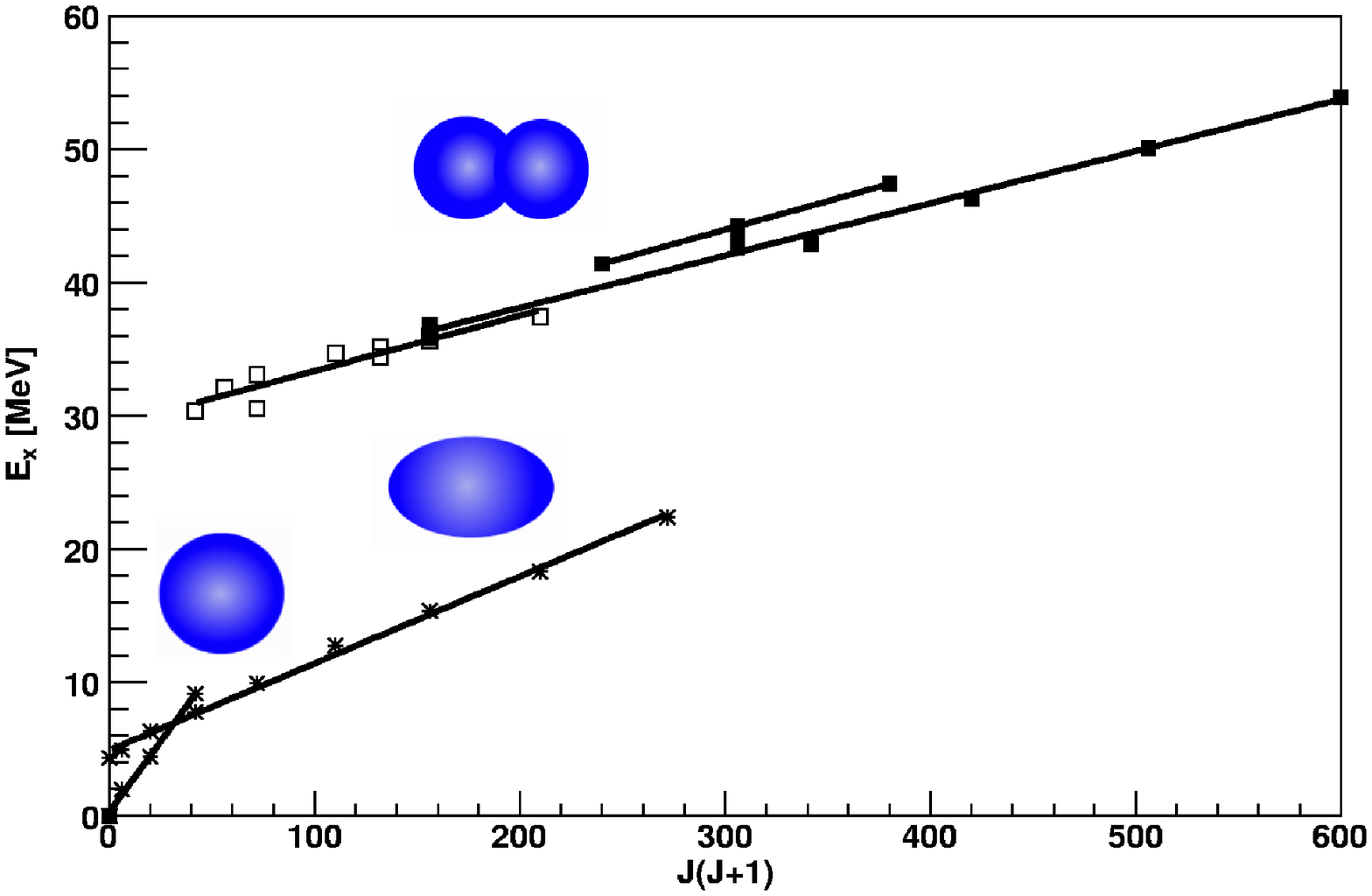}
\caption{\label{label}Rotational bands and deformed shapes in $^{36}$Ar. Excitation energies 
	of the ground state (spherical shape) and SD (ellipsoidal shape) bands~\cite{Svensson00}, respectively, and the energies of HD (dinuclear shape) band from 
	the quasimolecular resonances observed in the $^{12}$C+$^{24}$Mg 
	(open rectangles) \cite{Sciani09,Cindro79,Mermaz84,Pocanic85} and 
	$^{16}$O+$^{20}$Ne (full rectangles) \cite{Shimizu82,Gai84} reactions 
	are plotted as a function of J(J+1). This figure
	has been adapted from Refs.~\cite{Beck08a,Sciani09}.}
\end{minipage}\hspace{2pc}%
\begin{minipage}{18pc}
\includegraphics[width=18pc]{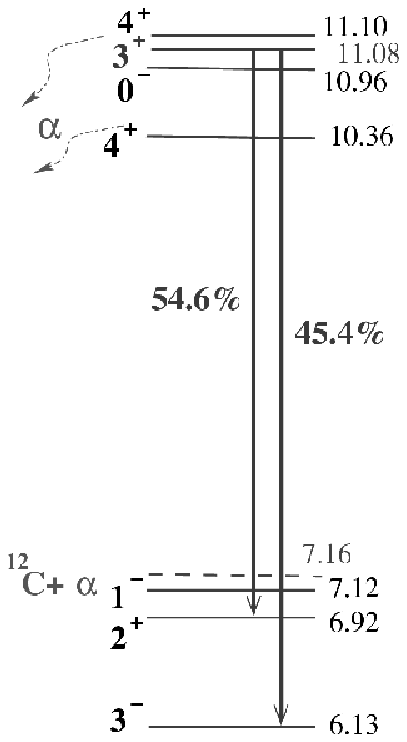}
\caption{\label{label}New partial (high-energy) level scheme of $^{16}$O 
corresponding to $\gamma$-ray transitions observed in the
$^{12}$C($^{24}$Mg,$^{20}$Ne)$^{16}$O$^{*}$ $\alpha$-transfer 
reactions. This figure has been adapted from Ref.~\cite{Beck09}.}
\end{minipage} 
\end{figure}

\newpage

There is a renewed interest in the spectroscopy of the $^{16}$O nucleus at high 
excitation energy \cite{Beck09}. Exclusive data were collected on $^{16}$O in the 
inverse kinematics reaction $^{24}$Mg$+^{12}$C studied at E$_{lab}$($^{24}$Mg) 
= 130 MeV with the BRS in coincidence with the {\sc Euroball IV} installed at 
the {\sc Vivitron} facility
\cite{Beck09}. From the $\alpha$-transfer reactions (both direct transfer
and deep-inelastic orbiting collisions \cite{Sanders99}), new information has been 
deduced on branching ratios of the decay of the 3$^{+}$ state of $^{16}$O at 
11.085~MeV $\pm$ 3 keV. The high-energy level scheme of $^{16}$O shown in
Fig.~3 indicates that this state does not $\alpha$-decay because of its non-natural parity 
(in contrast to the two neighbouring 4$^{+}$ states at 10.36~MeV and 11.10~MeV), 
but it $\gamma$ decays to the 2$^{+}$ state at 6.92~MeV (54.6 $\pm$ 2 $\%$) and 
to the 3$^-$ state at 6.13~MeV (45.4\%). 
By considering all the four possible transition types of the decay of the 3$^{+}$ 
state (\textit{i.e.} E$1$ and M$2$ for the 3$^{+}$ $\rightarrow$ 3$^{-}$ transition and, 
M$1$ and E$2$ for the 3$^{+}$ $\rightarrow$ 2$^{+}$ transition), our 
calculations yield the conclusion that $\Gamma_{3^+}<0.23$~eV, a value 
fifty times lower than known previously, which is an important result for the well studied $^{16}$O nucleus \cite{Beck09}.
Clustering effects in the light neutron-rich 
oxygen isotopes $^{17,18,19,20}$O will be discussed in Section 3.

In the study of the Bose-Einstein Condensation (BEC) the
$\alpha$-particle states in light $N$=$Z$ nuclei \cite{Tohsaki01,Oertzen10a,Yamada}, 
are of great importance. At present, the search for an experimental signature of 
BEC in $^{16}$O is of highest priority. 
A state with the structure of the ''Hoyle" state \cite{Hoyle54} in 
$^{12}$C coupled to an $\alpha$ particle is 
predicted in  $^{16}$O at about 15.1 MeV (the 0$^{+}_{6}$ state), the
energy of which is $\approx$ 700 keV above the 4$\alpha$-particle 
breakup threshold \cite{Funaki08}. However, any state in $^{16}$O equivalent 
to the so-called ''Hoyle" state \cite{Hoyle54} in $^{12}$C is most certainly 
going to decay by particle emission with very small, probably un-measurable, 
$\gamma$-decay branches, thus, very efficient particle-detection techniques will 
have to be used in the near future to search for them. BEC states are expected to 
decay by alpha emission to the ''Hoyle" state and could be found among the
resonances in $\alpha$-particle inelastic scattering on $^{12}$C decaying to 
that state (an early attempt to excite these states by $\alpha$ inelastic
scattering was presented in Ref.~\cite{Itoh04}), or could be observed in an $\alpha$-particle 
transfer channel leading to the 
$^{8}$Be--$^{8}$Be final state. Another possibility might be to perform Coulomb 
excitation measurements with intense $^{16}$O beams at intermediate energies.

\begin{figure}[h]
\begin{minipage}{18pc}
\includegraphics[width=18pc]{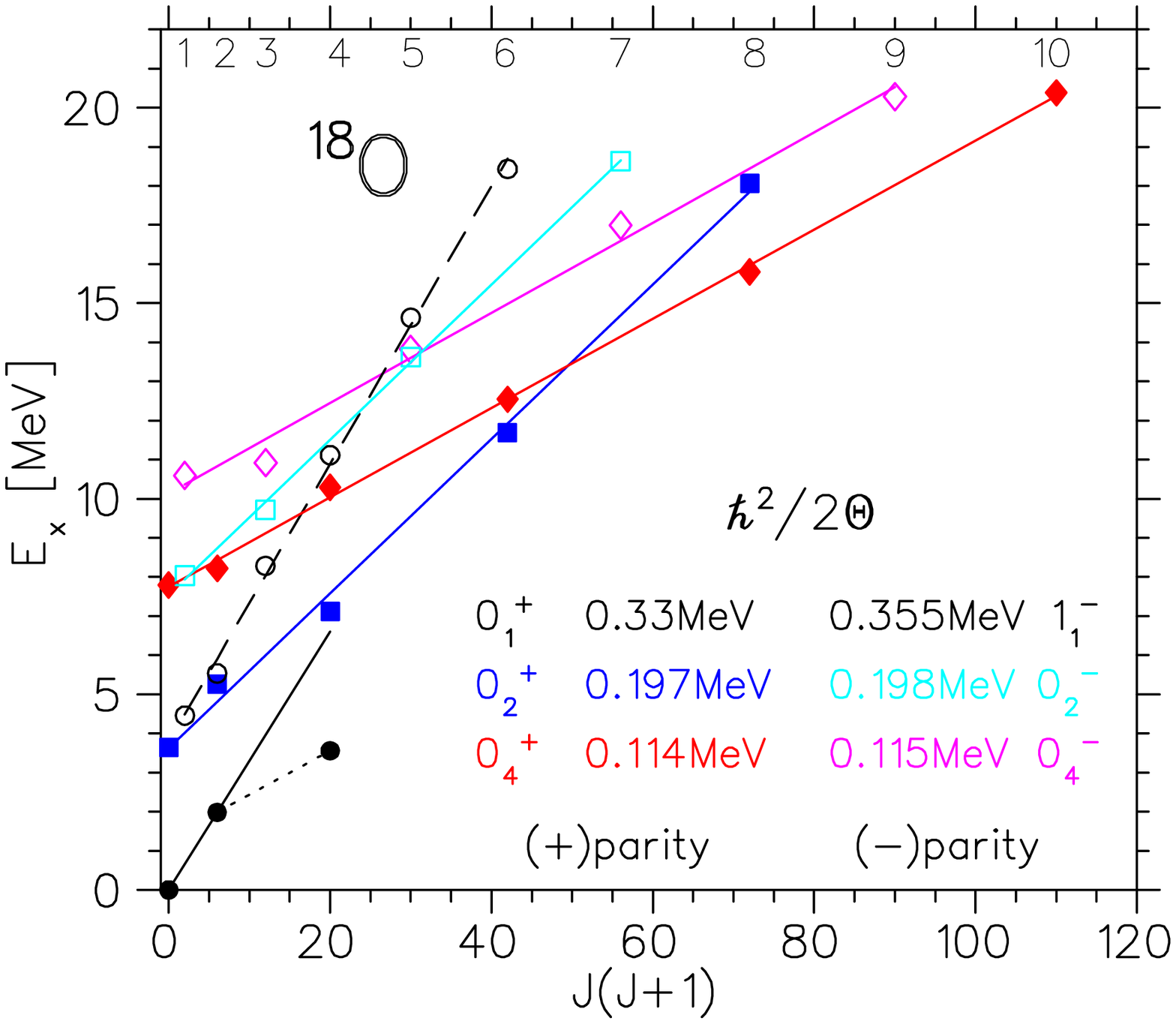}
\caption{\label{label}Overview of six rotational band structures observed in 
	$^{18}$O. Excitation energy systematics for the members of the rotational
	bands forming inversion doublets with K=0 are plotted as a function 
	of J(J+1). The curves are drawn to guide the eye for the slopes. The 
	indicated slope parameters contain information on
	the moments of inertia. Square symbols correspond to cluster bands,
	whereas diamonds symbols correspond to molecular bands. This figure
	is adapted from \cite{Oertzen10b}.}
\end{minipage}\hspace{2pc}%
\begin{minipage}{18pc}
\includegraphics[width=18pc]{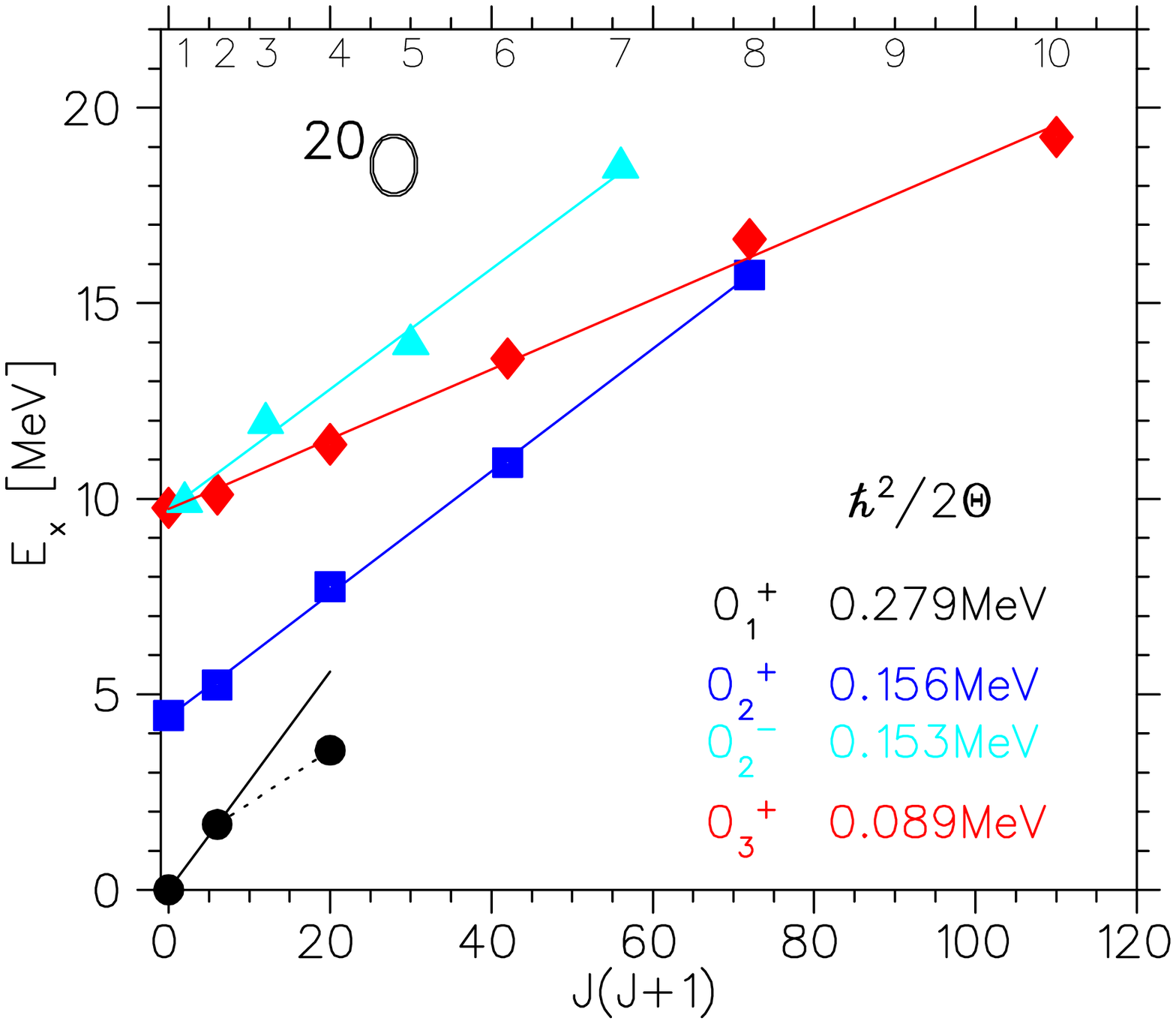}
\caption{\label{label}Overview of four rotational band structures observed in $^{20}$O.
	Excitation energy systematics for the members of the rotational
	bands forming inversion doublets with K=0 are plotted as a function 
	of J(J+1). The curves are drawn to guide the eye for the slopes.
	The indicated slope parameters contain information on
	the moments of inertia. Square and triangle symbols correspond to 
	cluster bands, whereas diamonds symbols correspond to molecular 
	bands.This figure is adapted from \cite{Oertzen09}}
\end{minipage} 
\end{figure}

\section{Clustering in light neutron-rich nuclei}
\label{sec:2}

As discussed previously, clustering is a general phenomenon observed also in 
nuclei with extra neutrons as it is presented in an extended "Ikeda"-diagram \cite{Ikeda} 
proposed by von Oertzen \cite{Oertzen01} (see the left panel of Fig.~1).
With additional neutrons, specific molecular structures 
appear with binding effects based on covalent molecular neutron orbitals. In 
these diagrams $\alpha$-clusters and $^{16}$O-clusters (as shown by the middle
panel of the diagram of Fig.~1) are the main ingredients. Actually, the $^{14}$C 
nucleus may play similar role in clusterization as the $^{16}$O one since it has similar  properties as a cluster: i) it has closed neutron p-shells, ii) first excited states are well above E$^{*}$ = 6 MeV, and iii) it has high binding energies for $\alpha$-particles.

A general picture of clustering and molecular configurations in light nuclei 
can be drawn from a detailed investigation of the light oxygen isotopes with
A $\geq$ 17. Here we will only present recent results on the even-even 
oxygen isotopes: $^{18}$O \cite{Oertzen10b} and $^{20}$O \cite{Oertzen09}. 
But very striking cluster states have also been found in odd-even oxygen 
isotopes such as: $^{17}$O \cite{Milin09} and $^{19}$O \cite{Oertzen11}. 

Fig.~4 gives an overview of all bands in $^{18}$O as a plot of excitation energies
as a function of J(J+1) together with their respective moments of inertia. In the
assignment of the bands both the dependence of excitation energies on J(J+1)
and the dependence of measured cross sections on 2J+1 \cite{Oertzen10b}
were considered. Slope parameters obtained in
a linear fit to the excitation energies \cite{Oertzen10b} indicate the moment
of inertia of the rotational bands given in Fig.~4. The intrinsic structure
of the cluster bands is reflection asymmetric, the parity projection gives an 
energy splitting between the partner bands.  
The assignment of the experimental molecular bands are supported by either
generator-coordinate-method \cite{Descouvemont} or Antisymmetrized Molecular
Dynamics (AMD) calculations \cite{Furutachi08}. 

\newpage

We can compare the bands of $^{20}$O \cite{Oertzen09} shown in Fig.~5 with those of $^{18}$O
displayed in Fig.~4. The first doublet (K=0$^{\pm}_{2}$) has a slightly larger moment of
inertia (smaller slope parameter) in $^{20}$O, which is consistent
with its interpretation as $^{14}$C--$^{6}$He or $^{16}$C--$^{4}$He 
molecular structures (they start well below the thresholds of 16.8 MeV and 
12.32 MeV, respectively). The second band, for which the negative parity 
partner is yet to be determined, has a slope parameter slightly smaller
than in $^{18}$O. This is consistent with the study of the bands in 
$^{20}$O by Furutachi et al. \cite{Furutachi08}, which clearly establishes 
parity inversion doublets predicted by AMD calculations for the 
$^{14}$C--$^6$He cluster and $^{14}$C-2n-$\alpha$ molecular structures.
The corresponding moments of inertia given in Fig.~4 and Fig.~5 are 
strongly suggesting large deformations for the cluster structures. We may
conclude that the  reduction of the moments of inertia of the lowest
bands of $^{18,20}$O is consistent with the assumption that the strongly bound $^{14}$C 
nucleus having equivalent properties to $^{16}$O, has a similar role
as $^{16}$O in relevant, less neutron rich nuclei. Therefore, the Ikeda-diagram 
\cite{Ikeda} and the "extended Ikeda-diagram" consisting of $^{16}$O cluster
cores with covalently bound neutrons \cite{Oertzen01} must be further extended to 
include also the $^{14}$C cluster cores as illustrated in Fig.~1. 

\newpage

\section{Summary, conclusions and outlook}

The connection of $\alpha$-clustering, quasimolecular resonances, orbiting 
phenomena and extreme deformations (SD, HD, ...) has been discussed in this 
work by using $\gamma$-ray spectroscopy of coincident binary fragments from 
either inelastic excitations and direct transfers (with small energy damping 
and spin transfer) or from orbiting (fully damped) processes \cite{Sanders99}
in the $^{24}$Mg+$^{12}$C reaction. From a careful analysis of the $^{16}$O+$^{20}$Ne 
$\alpha$-transfer exit-channel (strongly populated by orbiting) new information 
has been deduced on branching ratios of the decay of the 3$^{+}$ state of $^{16}$O 
at 11.089~MeV. This result is encouraging for a complete $\gamma$-ray spectroscopy 
of the $^{16}$O nucleus at high excitation energy. In addition, we have presented new results
on neutron-rich oxygen isotopes displaying very well defined molecular bands
in agreement with AMD predictions. Consequently, the extended Ikeda diagram
has been further extended for light neutron-rich nuclei by inclusion of the $^{14}$C 
cluster, similarly to the $^{16}$O one.  Of particular interest is the 
quest for the  4$\alpha$ states of $^{16}$O near the $^{8}$Be+$^{8}$Be and 
$^{12}$C+$\alpha$ decay thresholds, which correspond to the so-called ''Hoyle" state. 
The search for extremely 
elongated configurations (HD) in rapidly rotating medium-mass nuclei, which has 
been pursued by $\gamma$-ray spectroscopy measurements, will have to be 
performed in conjunction with charged-particle techniques in the near future 
since such states are most certainly  going to decay by particle emission (see \cite{Oertzen08,Wheldon08,Papka}).\\

\ack
We thank the {\sc Vivitron} staff and the {\sc Euroball} group of 
Strasbourg for their excellent support. C.B. would like to acknowledge 
Christian Caron (Springer) for 
initiating the new series of three volumes of \emph{Lecture Notes in Physics} dedicated 
to "Clusters in Nuclei".


\section*{References}
\begin{thebibliography}{99}

%\begin{thebibliography}{10}
%\providecommand{\url}[1]{{#1}}
%\providecommand{\urlprefix}{URL }
%\expandafter\ifx\csname urlstyle\endcsname\relax
 % \providecommand{\doi}[1]{DOI~\discretionary{}{}{}#1}\else
  %\providecommand{\doi}{DOI~\discretionary{}{}{}\begingroup
  %\urlstyle{rm}\Url}\fi

\bibitem{Greiner95} Greiner W, Park Y Jae, and Scheid W 1995 {\it Nuclear Molecules}, (Singapore: World 
Scientific)
\bibitem{Beck94} Beck C {\it et al.} 1994 \textit{Phys.\ Rev.\ C }{\bf 49} 2618
\bibitem{Freer07} Freer M 2007 \textit{Rep.\ Prog.\ Phys.} {\bf 70} 2149 
\bibitem{Horiuchi10} Horiuchi H 2010 Clusters in Nuclei - Vol.1, ed. Beck C, \textit{Lecture 
Notes in Physics} {\bf 818} 57; Horiuchi H, this conference.
\bibitem{Gupta10} Gupta R K 2010 Clusters in Nuclei - Vol.1, ed. Beck C, \textit{Lecture 
Notes in Physics} {\bf 818} 232
\bibitem{Oertzen06} von Oertzen W, Freer M, and  Kanada-En'yo Y 2006 \textit{Phys.\ Rep.\ }
{\bf 432} 43
\bibitem{Ikeda} Horiuchi H and Ikeda K 1968  \textit{Prog.\ Theor.\ Phys.} \bf 40 \rm
277
\bibitem{Oertzen01} von Oerzten W 2001 \textit{Eur.\ Phys.\ J.\ A} \bf 11 \rm  403
\bibitem{Milin} von Oerzten W and Milin M, 2013 Clusters in Nuclei - Vol.3,  ed. Beck C, \textit{Lecture Notes in Physics} (in press); Milin M, this
conference
\bibitem{Kanada10} Kanada-En'yo Y and Kimura M 2010 Clusters in Nuclei - Vol.1, ed. Beck C, \textit{Lecture Notes in Physics} {\bf 818} 129
\bibitem{Ikeda10} Ikeda K {\it et al.} 2010 Clusters in Nuclei - Vol.1, ed. C. Beck, \textit{Lecture Notes in Physics} {\bf 818} 165
\bibitem{Nakamura11} Nakamura N and Kondo Y 2012 Clusters in Nuclei - Vol.2, ed. Beck C, \textit{ Lecture Notes in Physics} {\bf 848} 67
\bibitem{Zagrebaev10} Zagrebaev V and Greiner W 2010 Clusters in Nuclei - Vol.1, ed.  Beck C, \textit{Lecture Notes in Physics} {\bf 818}  267 
\bibitem{Kamanin} Pyatkov Y and Kamanin D 2013 Clusters in Nuclei - Vol.3,  ed. Beck C, \textit{Lecture Notes in Physics} (in press)
\bibitem{Beck04a} Beck C 2004 {\it Nucl.\ Phys.} \bf A 738 \rm  24
\bibitem{Beck04b} Beck C 2004 {\it Int.\ J.\ Mod.\ Phys.} \bf E13 \rm  9
\bibitem{Cseh09} Cseh J {\it et al.} 2009 \textit{Phys.\ Rev.\ C }{\bf 80} 034320
\bibitem{Taniguchi09} Taniguchi Y {\it et al.} 2009 \textit{Phys.\ Rev.\ C} {\bf 80}, 
044316
\bibitem{Ichikawa11} Ichikawa T, Kanada-En'yo Y, and M\"oeller P,  2011
\textit{Phys.\ Rev.\ C} {\bf 83} 054319
\bibitem{Jenkins12} Jenkins D G {\it et al.}  2012 \textit{Phys.\ Rev.\ C} {\bf 86} 
064308
\bibitem{Jenkins13} Jenkins D G 2013 Clusters in Nuclei - Vol.3,  ed. Beck C, \textit{Lecture Notes in Physics} (in press); Jenskins D G this conference.
\bibitem{Darai12} Darai J, Cseh J and Jenkins D G 2012 \textit{Phys.\ Rev.\ C} {\bf 86}
064309
\bibitem{Kimura04} Kimura M and Horiuchi H 2004 \textit{Phys.\ Rev.\ C} {\bf 69} 
051304
\bibitem{Lonnroth10} L\"onnroth T {\it et al.} 2010 \textit{Eur.\ Phys.\ J.\ A} \bf 46 \rm  5 
\bibitem{Chandana10} Pandit D {\it et al.} 2010 \textit{Phys.\ Rev.\ C} {\bf 81} 061302 
\bibitem{Beck08a} Beck C {\it et al.} 2008 \textit{AIP Conf.\ Proc.} {\bf 1098} 207 
\bibitem{Svensson00} Svensson C E,  {\it et al.} 2000 \textit{Phys.\ Rev.\ Lett.} {\bf 85} 
2693
\bibitem{Sciani09} Sciani W {\it et al.} 2009 \textit{Phys.\ Rev.\ C} {\bf 80} 034319
\bibitem{Beck09} Beck C {\it et al.} 2009 \textit{Phys.\ Rev.\ C} {\bf 80} 034604
\bibitem{Ideguchi01} Ideguchi E {\it et al.} 2001 \textit{Phys.\ Rev.\ Lett.\ }\textbf {87} 
222501
\bibitem{Rousseau02} Rousseau M {\it et al.} 2002 \textit{Phys.\ Rev.\ C} \bf 66 \rm
034612
\bibitem{Taniguchi07} Taniguchi Y {\it et al.} 2007 \textit{Phys.\ Rev.\ C }\bf 76 \rm
044317  
\bibitem{Norrby10} Norrby M {\it et al.} (unpublished); Norrby M,
private communication and Norrby M, this conference.
\bibitem{Leary00} O'Leary C D {\it et al.} 2000 \textit{Phys. Rev. C} \bf 61 \rm  064314 
\bibitem{Fukada09} Fukada M {\it et al.} 2009 \textit{Phys.\ Rev.\ C} \bf 80 \rm  064613 
\bibitem{Salsac08} Salsac M D {\it et al.} 2008 \textit{Nucl. Phys.} {\bf A 801} 1 
\bibitem{Nouicer99} Nouicer M {\it et al.} 1999 \textit{Phys.\ Rev.\ C} \bf 60 \rm  
041303(R) 
\bibitem{Rudolph99} Rudolph D {\it et al.}  1999 \textit{Phys.\ Rev.\ Lett.} \bf 82 \rm  
3763
\bibitem{Beck01} Beck C {\it et al.} 2001 \textit{Phys.\ Rev.\ C} \bf 63 \rm  014607
\bibitem{Bhattacharya02} Bhattacharya C {\it et al.} 2002 \textit{Phys.\ Rev.\ C} \bf 65 \rm  
014611
\bibitem{Cindro79} Cindro N {\it et al.} 1979 \textit{Phys.\ Lett.} \bf 84B \rm  55
\bibitem{Mermaz84} Mermaz M C {\it et al.} 1984 \textit{Nucl.\ Phys.} \bf A 429 \rm  351
\bibitem{Pocanic85} Pocanic D and Cindro N 1985 \textit{Nucl.\ Phys.} \bf A 433 \rm  531 
\bibitem{Shimizu82} Schimizu J {\it et al.} 1982 \textit{Phys.\ Lett.} \bf 112B \rm  323 
\bibitem{Gai84} Gai M {\it et al.} 1984 \textit{Phys.\ Rev.\ C }\bf 30 \rm  925
\bibitem{Oertzen08} von Oertzen W {\it et al.} 2008 \textit{Eur.\ Phys.\ J.\ A} {\bf 36} 
279
\bibitem{Zagrebaev10b} Zagrebaev V I {\it et al} 2010  \textit{Phys.\ Rev.\ C} \bf 81 \rm  
044608
\bibitem{Pyatkov10} Pyatkov Y F {\it et al.} 2010 \textit{Eur.\ Phys.\ J.\ A} {\bf 45} 
29
\bibitem{Sanders99} Sanders S J, Szanto de Toledo A and
Beck C 1999 \textit{Phys.\ Rep.} \bf 311 \rm 487 
\bibitem{Tohsaki01} Tohsaki A, \textit{et al.} 2001 \textit{Phys. Rev. Lett.} \bf 87 \rm   192501 
\bibitem{Oertzen10a} von Oertzen W 2010 Clusters in Nuclei - Vol.1, ed. Beck C, \textit{Lecture 
Notes in Physics} {\bf 818} 102
\bibitem{Yamada} Yamada T {\it et al.} 2012 Clusters in Nuclei - Vol.2, ed. Beck C, \textit{ Lecture Notes in Physics} {\bf 848} 229
\bibitem{Hoyle54}Hoyle F 1954 \textit{Astrophys.\ J.\ Suppl.\ Ser.} \bf 1 \rm  121 
\bibitem{Funaki08}Funaki Y {\it et al.} 2008 \textit{Phys.\ Rev.\ Lett.} \bf 101 \rm
082502
\bibitem{Itoh04} Itoh M {\it et al.} 2004 \textit{Nucl.\ Phys.\ A} \bf738 \rm  268
\bibitem{Oertzen10b} von Oertzen W {\it et al.} 2010 \textit{Eur.\ Phys.\ J.\ A} \bf 43 \rm
17
\bibitem{Oertzen09} von Oertzen W {\it et al.} 2009 \textit{AIP Conf.\ 
Proc.} \bf 1165 \rm  19
\bibitem{Milin09} Milin M {\it et al.} 2009 \textit{Eur. Phys. J. A} \bf 41 \rm   335
\bibitem{Oertzen11} Bohlen H G {\it et al.} 2011
\textit{Eur.\ Phys.\ J.\ A} \bf 47 \rm 44
\bibitem{Descouvemont} Descouvemont P and Baye D  1985 \textit{Phys.\ Rev.\ C} \bf 31 \rm
2274
%; P. Descouvemont, Phys. Lett. \bf 437B \rm, 7 (1998).
\bibitem{Furutachi08} Furutachi N {\it et al.} 2008 \textit{Prog.\ Theor.\ Phys.} 
\bf 119 \rm  403
\bibitem{Wheldon08} Wheldon C {\it et al.} 2008 \textit{Nucl. Phys. A} \bf 811 \rm  276
\bibitem{Papka} Papka P and Beck C 2012 Clusters in Nuclei - Vol.2 ed. Beck C, \textit{Lecture 
Notes in Physics} {\bf 848} 299
\end{thebibliography}
\end{document}